\documentstyle[epsfig]{aipproc}

\begin{document}
%%%%%%%%%%%%%%%%%%%%%%%%%%%%%%%%%%%%%%%%%%%%%%%%%%%%%%%%%%%%%%%%%%%%%%

\pagestyle{empty}

\begin{flushleft}
\large
SAGA-HE-169-00 \hfill December 7, 2000 \\
RIKEN-AF-NP-378 \\
TMU-NT-00-1 \\
\end{flushleft}

\vspace{1.2cm}

\begin{center}

\LARGE\bf 
Determination of \\
polarized parton distribution functions \\

\vspace{1.2cm}

\Large
M. Hirai$^*$, H. Kobayashi$^{\dagger}$, 
and \underline{M. Miyama}$^{\ddagger}$ \\
(Asymmetry Analysis Collaboration)  \\

\vspace{0.6cm}

\large
$^*$Department of Physics, Saga University, Saga 840-8502, Japan \\
\vspace{0.1cm}
$^{\dagger}$RIKEN BNL Research Center, Upton, NY 11973, U.S.A. \\
\vspace{0.1cm}
$^{\ddagger}$Department of Physics, Tokyo Metropolitan University, \\
Tokyo 192-0397, Japan \\
 
\vspace{1.0cm}
 
\LARGE
Talk given at the 14th International \\
Spin Physics Symposium \\
Osaka, Japan, October 16 -- 21, 2000 \\
(talk on October 20, 2000) \\
 
\end{center}
 
\vfill
\noindent
{\rule{6.0cm}{0.1mm}} \\
 
\vspace{-0.3cm}
\normalsize
\noindent
$^*$98td25@edu.cc.saga-u.ac.jp,
$^{\dagger}$hyuki@bnl.gov,
$^{\ddagger}$miyama@comp.metro-u.ac.jp, \\ 
\ Information on their research is available 
at http://spin.riken.bnl.gov/aac/. \\

\vspace{+0.1cm}
\hfill
{\large to be published in proceedings}

\vfill\eject
\setcounter{page}{1}
\pagestyle{plain}

%%%%%%%%%%%%%%%%%%%%%%%%%%%%%%%%%%%%%%%%%%%%%%%%%%%%%%%%%%%%%%%%%%%%%%

\title{Determination of polarized parton distribution functions}

\author{
M. Hirai \thanks{98td25@edu.cc.saga-u.ac.jp},
H. Kobayashi \thanks{hyuki@bnl.gov},
and \underline{M. Miyama}
\thanks{miyama@comp.metro-u.ac.jp} \\
(Asymmetry Analysis Collaboration)}
\address{
$^1$Department of Physics, Saga University, Saga 840-8502, Japan\\
$^2$RIKEN BNL Research Center, Upton, NY 11973-5000,
U.S.A.\\
$^3$Department of Physics, Tokyo Metropolitan University,
Tokyo 192-0397, Japan
}

\maketitle

%%%%%%%%%%%%%%%%%%%%%%%%%%%%%%%%%%%%%%%%%%%%%%%%%%%%%%%%%%%%%%%%%%%%%%
\begin{abstract}
We study parametrization of polarized parton distribution
functions in the $\alpha_s$ leading order (LO) and
in the next-to-leading order (NLO). From $\chi^2$ fitting
to the experimental data on $A_1$,
optimum polarized distribution functions are determined.
The quark spin content $\Delta\Sigma$ is very sensitive
to the small-$x$ behavior of antiquark distributions which
suggests that small-$x$ data are needed for precise
determination of $\Delta\Sigma$. We propose three sets of
distributions and also provide FORTRAN library for our
distributions.

\end{abstract}

%%%%%%%%%%%%%%%%%%%%%%%%%%%%%%%%%%%%%%%%%%%%%%%%%%%%%%%%%%%%%%%%%%%%%%
\section*{Introduction}

Experimental data on polarized structure functions $g_1$
have been accumulated for the last several years. The data
with the proton, deuteron, and $^3$He targets are now available
and new data are expected to be given by RHIC at BNL,
COMPASS at CERN, and other facilities in the near future.
In the light of such progresses and future projects,
we should summarize present knowledge of polarized parton
distribution functions by using all available data at this
stage. For this purpose, we formed the group called Asymmetry
Analysis Collaboration (AAC) and tried to determine the polarized
parton distribution functions.

In this paper, we explain our analysis to determine
optimum polarized distributions by $\chi^2$ fitting
to the experimental data of spin asymmetry $A_1$.
In addition to the discussions based on the work
in Ref.\cite{MM:AAC00}, we introduce a FORTRAN library
for our distributions as a recent progress.

%%%%%%%%%%%%%%%%%%%%%%%%%%%%%%%%%%%%%%%%%%%%%%%%%%%%%%%%%%%%%%%%%%%%%%
\section*{Parametrization}

We determine initial polarized parton distributions at
$Q^2$ = 1.0 GeV$^2$ ($\equiv Q^2_0$). Considering the
counting rule, we adopt the following functional form:
\begin{equation}
\Delta f_i(x, Q^2_0) = A_i \, x^{\alpha_i} \,
               (1 + \gamma_i \, x^{\lambda_i}) \, f_i(x, Q^2_0) .
\label{MM:eq1}
\end{equation}
Here, $\Delta f_i$ and $f_i$ represent the polarized and
unpolarized parton distributions, respectively and
$A_i$, $\alpha_i$, $\gamma_i$, and $\lambda_i$ are
free parameters. These parameters are constrained by the
positivity condition, $| \,\Delta f_i (x) \, | \le f_i (x)$.
We simply apply it not only for the leading-order (LO) case
but also for the next-to-leading-order (NLO) case
although it is valid only in LO.
Furthermore, we assume SU(3) flavor-symmetric sea
because we don't have enough data to extract flavor
dependence of polarized sea.
Under this assumption, the first moments of $\Delta u_v$
and $\Delta d_v$ can be fixed by the axial charges
for octet baryon, F and D.
Therefore, we should determine $\Delta u_v$, $\Delta d_v$,
$\Delta \bar q$, and $\Delta g$ distributions
and total number of the parameters becomes 14.

As the experimental data to which the parameters are fitted,
we chose spin asymmetry $A_1$ instead of $g_1$
since $A_1$ is closer to direct observable
in experiment rather than $g_1$. $A_1$ is given by
\begin{equation}
A_1(x,Q^2) \simeq \frac{g_{1}(x,Q^2)}{F_{1}(x,Q^2)}
           = g_{1}(x,Q^2) \frac{2x [1+R(x,Q^2)]}{F_{2}(x,Q^2)} ,
\end{equation}
where, $R$ is the ratio of longitudinal to transverse
cross-sections. In our analysis, $A_1$ is calculated by
using $F_2$ obtained by GRV98 distributions, $g_1$
obtained by our parametrized distributions,
and $R$ taken from the SLAC-1990 analysis. Then,
$\chi^2=\sum[A_1^{\rm data}(x,Q^2)-A_1^{\rm calc}(x,Q^2)]^2/
[\sigma^{data}(x,Q^2)]^2$ is calculated and is minimized
by the subroutine MINUIT. Here, $A_1^{data}$, $A_1^{calc}$,
and $\sigma^{data}$ indicate the experimental $A_1$,
calculated $A_1$, and experimental error, respectively.

%%%%%%%%%%%%%%%%%%%%%%%%%%%%%%%%%%%%%%%%%%%%%%%%%%%%%%%%%%%%%%%%%%%%%%
\section*{Results and discussions}

\begin{figure}[t!] % fig 1
\centerline{\epsfig{file=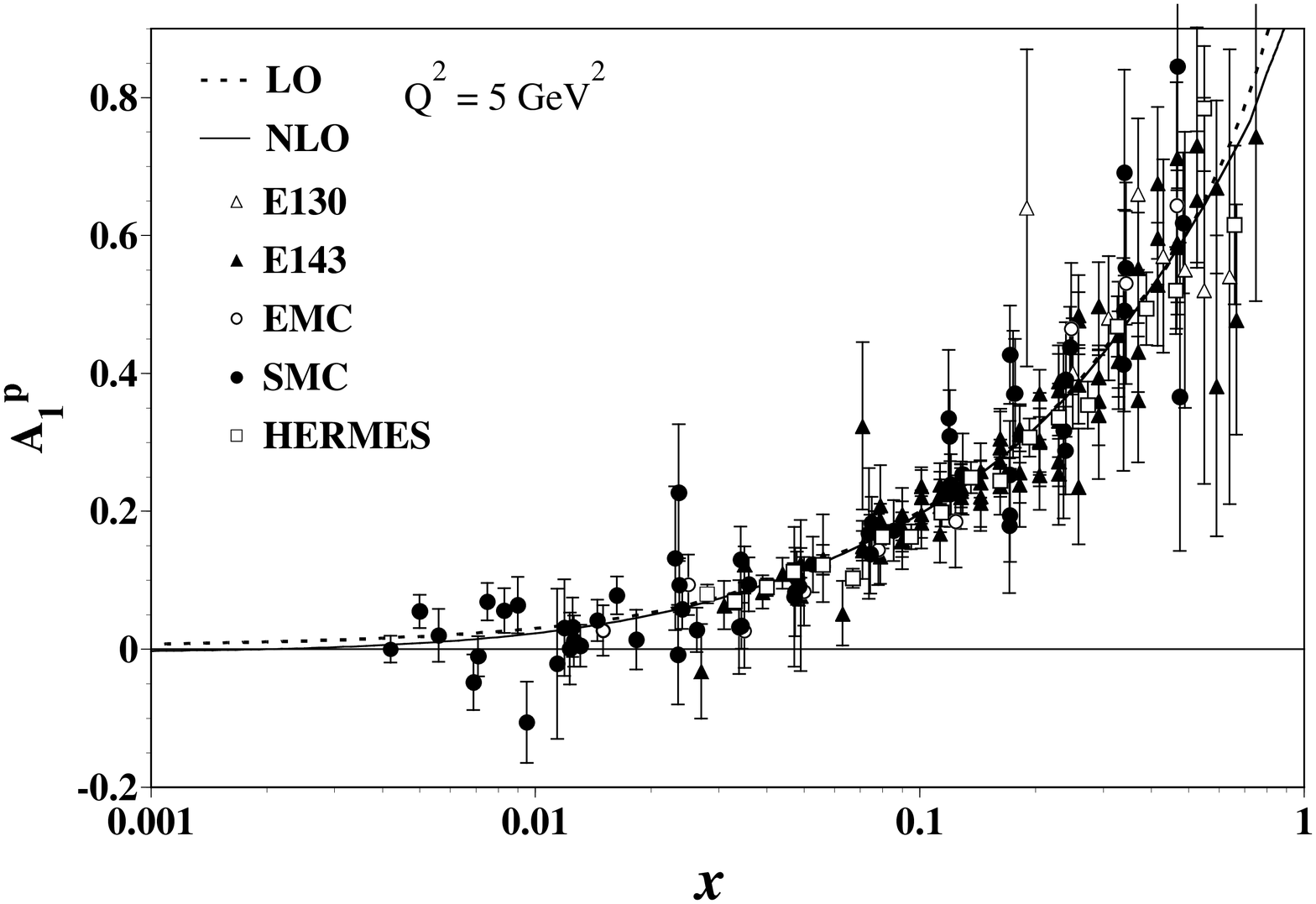,width=6.0cm}
\hspace{1.0cm}\epsfig{file=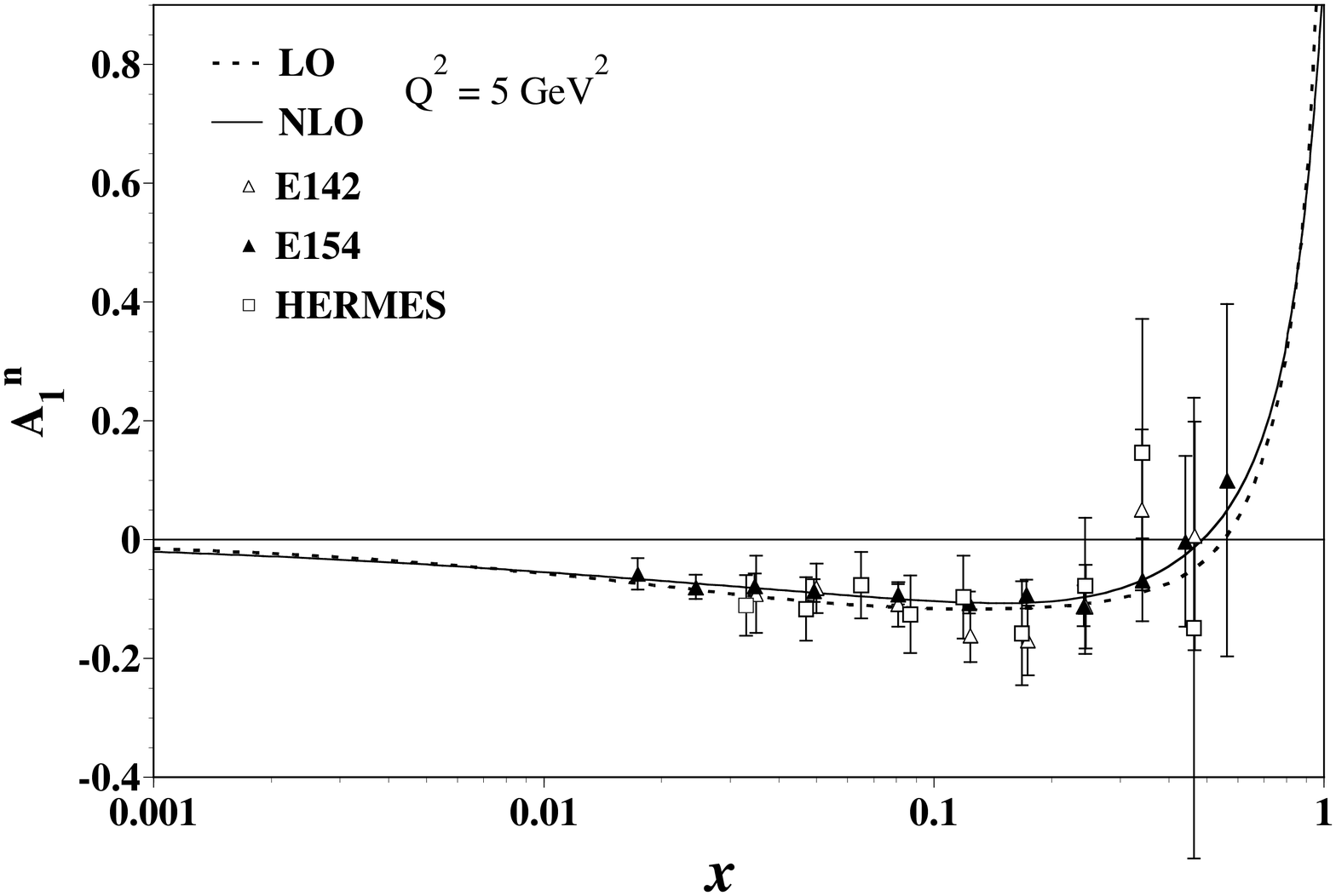,width=6.0cm}}
\vspace{10pt}
\caption{Spin asymmetries $A_1$ for the proton (left figure)
and neutron (right figure).}
\label{MM:fig1}
\end{figure}

\begin{figure}[b!] % fig 2
\centerline{\epsfig{file=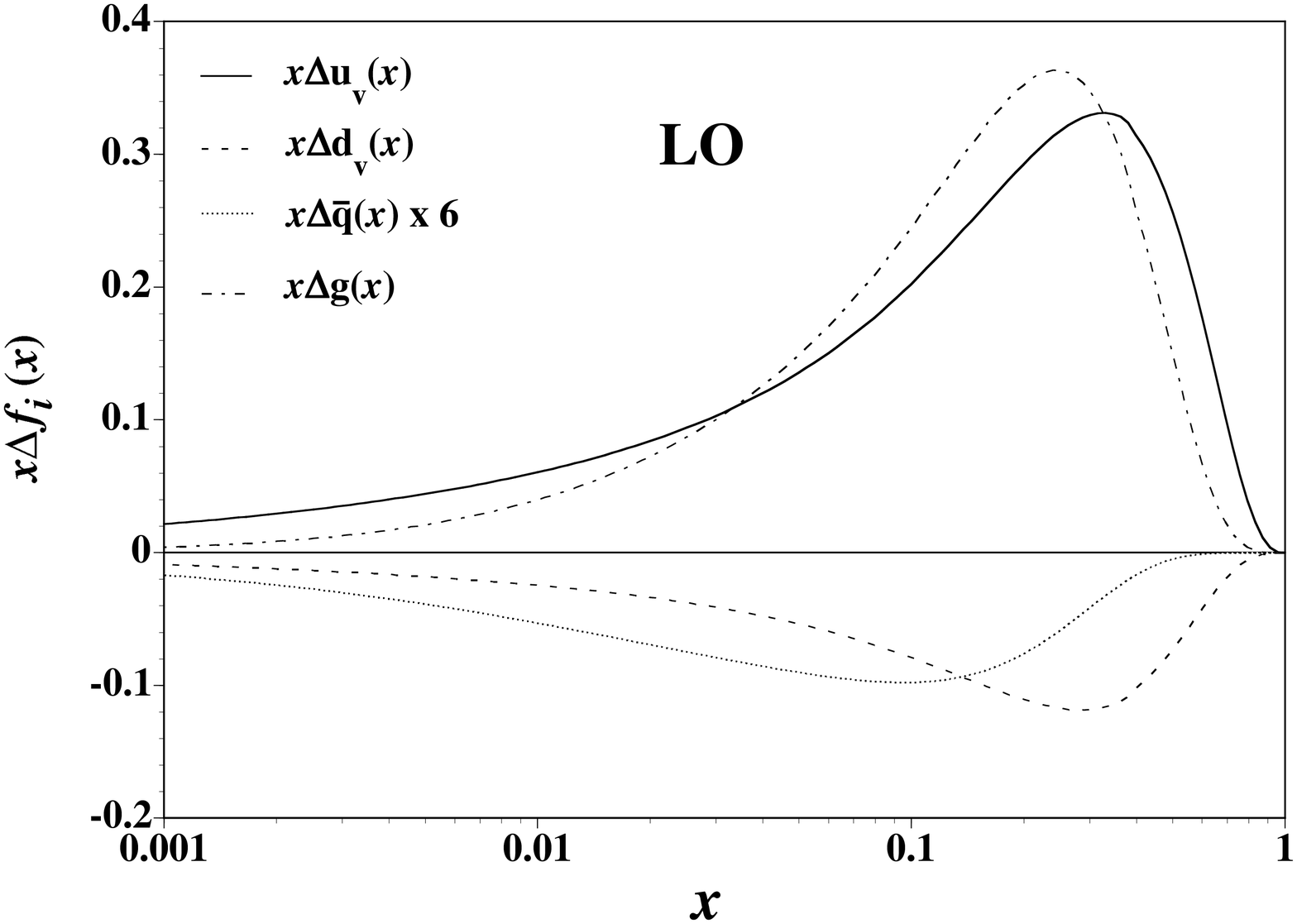,width=6.0cm}
\hspace{1.0cm}\epsfig{file=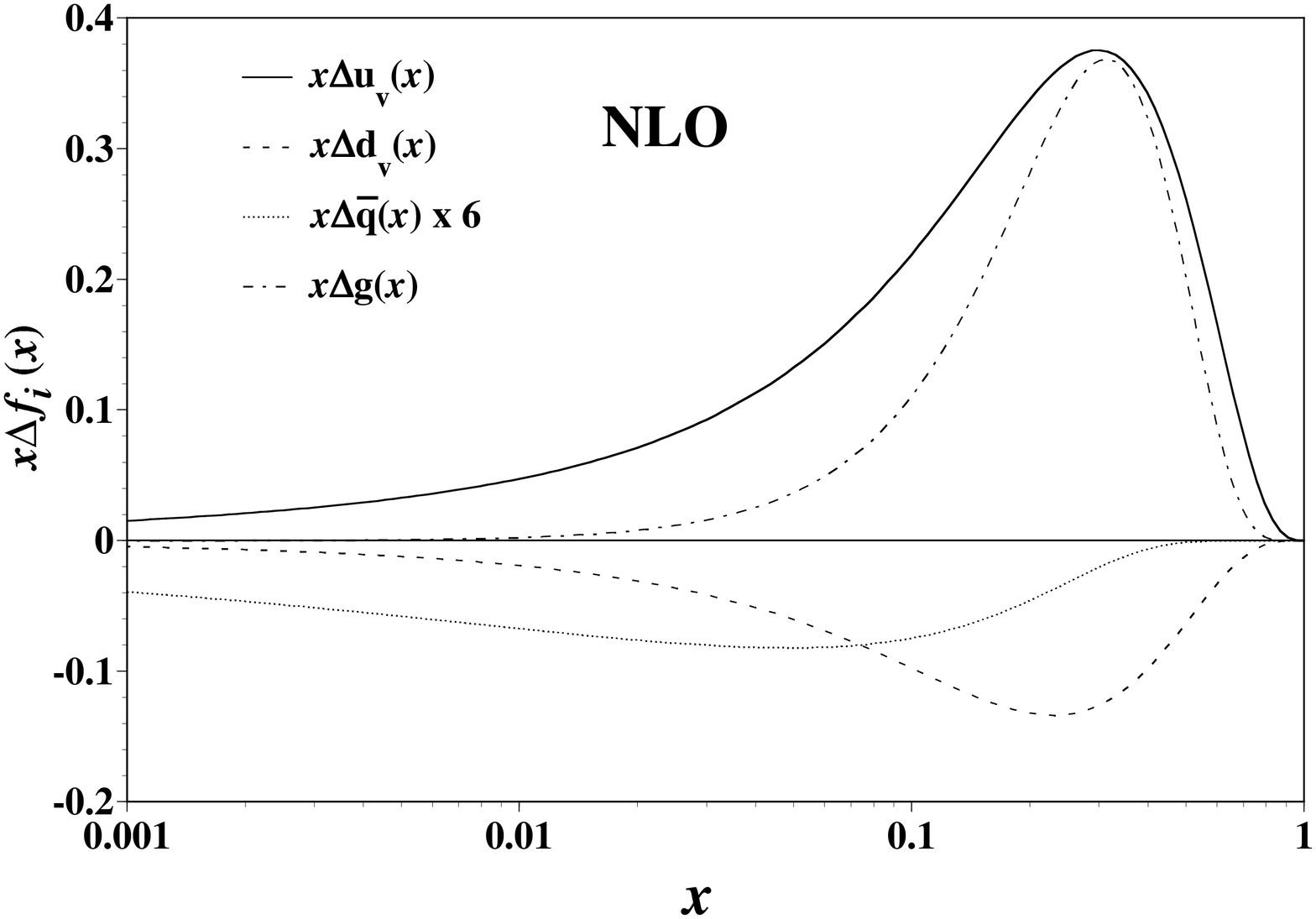,width=6.0cm}}
\vspace{10pt}
\caption{Obtained polarized distributions at $Q^2$
= 1.0 GeV$^2$.}
\label{MM:fig2}
\end{figure}

From the $\chi^2$ analysis, we obtain the results
with $\chi^2$ = 322.6 in the LO case, and
$\chi^2$ = 300.4 in the NLO case for 375 data points.
The NLO $\chi^2$ is significantly smaller than the LO one.
It suggests that the NLO analysis is necessary
for precise analysis.
The obtained $A_1$ for the proton and for the neutron
at $Q^2$ = 5.0 GeV$^2$ are shown in Fig.\ref{MM:fig1}
with the experimental data. Although the comparison
is not straightforward because the data are taken
at various $Q^2$, the obtained parameters reproduce
the data well both in the LO and the NLO case.
Figure \ref{MM:fig2} show the obtained polarized parton
distribution functions at $Q^2$ = 1.0 GeV$^2$.
In our analysis, the first moment of $\Delta u_v$
($\Delta d_v$) is fixed by positive (negative) value
and the obtained distribution becomes positive (negative).
Furthermore, antiquark and gluon distributions become
negative and positive, respectivly.

%---------------------------------------------------------------
%---------------------------------------------------------------

%---------------------------------------------------------------
%---------------------------------------------------------------

For these distributions, quark spin content becomes
$\Delta\Sigma$ = 0.201 for LO and $\Delta\Sigma = 0.051$
for NLO at $Q^2$=1.0 GeV$^2$.
The NLO $\Delta\Sigma$ is significantly smaller than
the LO one. It is also smaller than the generally
quoted values in other analyses that range
in 0.1 $\sim$ 0.3.
The difference of $\Delta\Sigma$ comes mainly from
the difference of $\Delta\bar q$ in the small-$x$ region.
Because we don't have a small-$x$ data at this stage,
we should fix the small-$x$ behavior of $\Delta\bar q$
from theoretical suggestions.
We perform $\chi^2$ fitting by fixing the parameter
$\alpha_{\bar q}$, which controls the small-$x$ behavior
of $\Delta\bar q$, to $\alpha_{\bar q}$ = 1.0 and 1.6 
according to the predictions 
of the Regge theory and the perturbative QCD.
These values are rather larger than the value
$\alpha_{\bar q}$ = 0.32 $\pm$ 0.22 which is obtained
by the previous NLO analysis.
As a result, we get $\Delta\Sigma$ = 0.241
($\chi^2$ = 305.8) for $\alpha_{\bar q}$ = 1.0
and $\Delta\Sigma$ = 0.276 ($\chi^2$ = 323.5)
for $\alpha_{\bar q}$ = 1.6. The obtained $\Delta\Sigma$
are larger than the previous NLO result and are within the
usually quoted range. From these results, we find that
$\Delta\Sigma$ is very sensitive to the small-$x$
behavior of the antiquark distributions.
In order to determine the quark spin content precisely,
the data in the small-$x$ region are needed.
Because $\chi^2$ for the $\alpha_{\bar q}$ = 1.0 case
is close to the one for the previous NLO,
this set can be also considered as a good parametrization.
Therefore, we propose three sets of distributions,
LO, NLO with free $\alpha_{\bar q}$ (NLO-1),
and NLO with $\alpha_{\bar q}$ = 1.0 (NLO-2),
as AAC polarized distribution functions.

Finally, we briefly comment on the error estimate of
the obtained distributions \cite{MM:ERROR}.
The uncertainties of our distributions can be
estimated by using the error matrix which is obtained
by MINUIT. The 1-$\sigma$ boundary of
the distribution $F$ is given by

\begin{equation}
(\delta F)^2 = \sum_i \sum_j \frac{\partial F}{\partial a_i}
               V_{ij} \frac{\partial F}{\partial a_j} ,
\end{equation}

\noindent
where, $V_{ij}$ is the error matrix element
for parameters $a_i$ and $a_j$.
This analysis is in progress and will be reported elsewhere.

%%%%%%%%%%%%%%%%%%%%%%%%%%%%%%%%%%%%%%%%%%%%%%%%%%%%%%%%%%%%%%%%%%%%%%
\section*{Library}

For practical application of our distributions,
we provide a library program which is given as
FORTRAN subroutine AACPDF(ISET, Q2, X, POLPDF, STRUCT).
The users should call this subroutine in their programs
with the input parameters ISET, Q2, and X.
This library contains three sets of distributions,
LO, NLO-1, and NLO-2, and the parameter ISET designates
which set is used. When ISET=1, 2, or 3, the returned values
are those in the LO, NLO-1, or NLO-2, respectively.
The parameters Q2 and X specify $Q^2$ and Bjorken-$x$
at which the distributions are calculated.
The allowed ranges are 1.0 GeV$^2$ $\leq Q^2 \leq$
10$^6$ GeV$^2$ and 10$^{-9}$ $\leq x \leq$ 1.0.
The obtained values are returned by the arrays POLPDF($-$3:3)
and STRUCT(3). The available polarized parton distribution
functions and the structure functions are listed
in Table \ref{MM:tab1}.
It should be noted that the returned values are the
distributions and the structure functions multiplied by $x$.

\begin{table}[b!]
\caption{Available distributions and structure functions.}
\label{MM:tab1}
\begin{tabular}{c|l|l}
\multicolumn{1}{c|}{I} & \multicolumn{1}{c|}{POLPDF(I)}
& \multicolumn{1}{c}{STRUCT(I)} \\
\tableline
-3 & $\bar s$ quark 
[$x\Delta\bar s(x, Q^2)$ = $x\Delta s(x, Q^2)$)]
& \multicolumn{1}{c}{---} \\
-2 & $\bar d$ quark 
[$x\Delta\bar d(x, Q^2)$ = $x\Delta d_{sea}(x, Q^2)$]
& \multicolumn{1}{c}{---} \\
-1 & $\bar u$ quark 
[$x\Delta\bar u(x, Q^2)$ = $x\Delta u_{sea}(x, Q^2)$]
& \multicolumn{1}{c}{---} \\
\ 0 & gluon [$x\Delta g(x, Q^2)$]
& \multicolumn{1}{c}{---} \\
\ 1 & u-valence quark [$x\Delta u_v(x, Q^2)$]
& $x g_1(x, Q^2)$ for proton \\
\ 2 & d-valence quark [$x\Delta d_v(x, Q^2)$]
& $x g_1(x, Q^2)$ for neutron\\
\ 3 & s quark [$x\Delta s(x, Q^2)$ = $x\Delta\bar s(x, Q^2)$]
& $x g_1(x, Q^2)$ for deuteron \\
\end{tabular}
\end{table}

The distribution values are obtained by interpolating
the grid data which are provided as DATA statements
in the subroutine. The number of the grid points is 23
for the variable $Q^2$ and 68 for the variable $x$.
Because the $Q^2$ dependence of the distributions is
almost linear function of $t \equiv \ln Q^2$, we simply use
the linear interpolation of $t$ for the variable $Q^2$.
If $t'$ is in the range $t_i \leq t' < t_{i+1}$,
the distributions at $t = t'$ are approximated by

\begin{equation}
f(t') \approx \frac{t_{i+1} - t'}{t_{i+1} - t_i} f(t_i)
      + \frac{t' - t_i}{t_{i+1} - t_i} f(t_{i+1}) .
\end{equation}

\noindent
On the other hand, the $x$ dependence of the distributions
is rather complicated. Therefore, we use the cubic spline
interpolation for the variable $x$.
If $x'$ is in the range $x_i \leq x' < x_{i+1}$,
the distributions at $x = x'$ are approximated by

\begin{equation}
f(x') \approx f(x_i) + B_i(x' - x_i) + C_i(x' - x_i)^2
      + D_i(x' - x_i)^3 .
\end{equation}

\noindent
Here, $B_i$, $C_i$, $D_i$ are the spline coefficients
which can be calculated by the grid data.
By using these interpolation methods and the grid data,
the AACPDF returns the values of the polarized parton
distribution functions and the structure functions $g_1$
at the specified $Q^2$ and $x$ point.
We verified that the distributions obtained by the library
reproduce well the results in Ref.\cite{MM:AAC00}.

AAC library can be downloaded from our web page
\cite{MM:AACWEB}. The subroutine AACPDF is in the file
`` aac.f ". This file also includes the subroutine SPLINE
and the function ISERCH which are used in the AACPDF.

%%%%%%%%%%%%%%%%%%%%%%%%%%%%%%%%%%%%%%%%%%%%%%%%%%%%%%%%%%%%%%%%%%%%%%
\section*{Summary}

The polarized parton distribution functions have been
determined by $\chi^2$ fitting to $A_1$ experimental data.
As a result, we found that the NLO $\chi^2$ is significantly
smaller than the LO one. It implies that the NLO analysis is
important. In the NLO analysis, the obtained $\Delta\Sigma$
is rather smaller than the usually quoted values,
and the differences come from the small-$x$ behavior
of $\Delta\bar q$. In order to fix $\Delta\Sigma$ precisely,
the small-$x$ measurements are required. We propose
three sets of AAC distributions, LO, NLO-1
($\alpha_{\bar q}$ : free), and NLO-2
($\alpha_{\bar q}$ = 1.0 fixed).
The FORTRAN library is provided for calculating
the AAC distributions numerically.
This library is useful for practical application of the AAC
distributions and available at our web site
\cite{MM:AACWEB}.

%%%%%%%%%%%%%%%%%%%%%%%%%%%%%%%%%%%%%%%%%%%%%%%%%%%%%%%%%%%%%%%%%%%%%%
\section*{Acknowledgements}

M.H. and M.M. were supported by the JSPS Research Fellowships
for Young Scientists and by the Grant-in-Aid
from the Japanese Ministry of Education, Science, and Culture.
This talk is based on the work with Y.~Goto, N.~Hayashi,
H.~Horikawa, S.~Kumano, T.~Morii, N.~Saito, T.-A.~Shibata,
E.~Taniguchi, and T.~Yamanishi.

%%%%%%%%%%%%%%%%%%%%%%%%%%%%%%%%%%%%%%%%%%%%%%%%%%%%%%%%%%%%%%%%%%%%%%

\end{document}